\documentclass[aps,prb,preprint]{revtex4}
\usepackage{graphicx}
\usepackage{amsmath}
\usepackage{dcolumn}
\usepackage{bm}
\usepackage{multirow}
\usepackage{threeparttable}
\usepackage{xcolor}

\begin{document}

\title{Insight into the description of van der Waals forces for benzene adsorption on transition metal (111) surfaces }
\author{Javier Carrasco$^1$}
\email{jcarrasco@cicenergigune.com}
\author{Wei Liu$^2$}
\author{Angelos Michaelides$^3$}
\author{Alexandre Tkatchenko$^2$}
\email{tkatchenko@fhi-berlin.mpg.de}
\affiliation{$^1$CIC Energigune, Albert Einstein 48, 01510 Mi\~nano, \'Alava, Spain \\
$^2$Fritz-Haber-Institut der Max-Planck-Gesellschaft, Faradayweg 4-6, D-14195, Berlin, Germany\\
$^3$Thomas Young Centre, London Centre for Nanotechnology and Department of Chemistry, University College London, London WC1E 6BT, United Kingdom}


\begin{abstract}
Exploring the role of van der Waals (vdW) forces on the adsorption of molecules on extended metal surfaces  has 
become possible in recent years thanks to exciting developments in density functional theory (DFT). Among these 
newly developed vdW-inclusive methods, interatomic vdW approaches that account for the nonlocal screening 
within the bulk
[V.\ G.\ Ruiz, W.\ Liu, E.\ Zojer, M.\ Scheffler, and A.\ Tkatchenko, Phys. Rev. Lett. {\bf 108}, 146103 (2012)]
and improved nonlocal functionals
[J.\ Klime{\v{s}}, D.\ R.\  Bowler, and A.\ Michaelides, J. Phys.: Condens. Matter {\bf 22}, 022201(2010)]
have emerged as promising candidates to account efficiently 
and accurately for the lack of long-range vdW forces in most popular DFT exchange-correlation 
functionals. Here we have used these two approaches to compute  benzene adsorption on a range of close-packed (111) surfaces upon which 
it  either physisorbs  (Cu, Ag, and Au) or chemisorbs (Rh, Pd, Ir, and Pt). We have thoroughly 
compared the performance between the two classes of vdW-inclusive methods and when available compared the results obtained with
experimental data. By examining 
the computed adsorption energies, equilibrium distances, and binding curves we conclude that both methods 
allow for an accurate treatment of adsorption at equilibrium adsorbate-substrate distances. To this end, explicit inclusion of electrodynamic screening
in the interatomic vdW scheme and optimized exchange functionals in the case of nonlocal vdW density functionals is
mandatory. Nevertheless, some discrepancies are found between these two classes of methods at large
adsorbate-substrate separations.
\end{abstract}

\maketitle

\section{INTRODUCTION}

Van der Waals (vdW) forces are ubiquitous in the binding of atoms and molecules. Although they are relatively
weak compared to, for example, covalent and ionic bonding, vdW forces play an important role in  fields
as diverse as macromolecular biochemistry,
supramolecular chemistry, and condensed matter physics.~\cite{stone1996,parsegian2005,french2010}
Yet despite this
importance our current understanding of vdW forces comes mainly from the study of atoms and small molecules
in the gas phase, and much less is known about vdW interactions in large aggregates and extended systems of 
interest in basic and applied science. It is not surprising then that accurately accounting for vdW forces and 
understanding the role they play in extended systems has become 
a thriving topic of research in recent years.

From a theoretical viewpoint, density functional theory (DFT) is the method of choice
to 
gain insight into
the electronic structure of relatively large systems (typically hundreds
of atoms). However the widely used generalized gradient approximation (GGA) functionals, which are generally employed in 
DFT studies of extended systems, fail to describe non-local vdW forces (as reported, for example, in Ref.[\cite{lee2011}]).
Fortunately, great progress has been made recently in remedying this long-standig problem (see,
\emph{e.g.}, Refs. [\cite{dion2004,tkatchenko2009,grimme2010,klimes2010}] and Ref. [\cite{klimes2012}] for a recent review),
making it possible to account
efficiently and accurately for the long-range vdW energy of solids, extended surfaces, and adsorption
processes.~\cite{tkatchenko2010,victor2012,javi2011} Weakly interacting atoms and molecules on metal
surfaces have become the workhorse systems for understanding how vdW forces influence the interaction
of adsorbates with solid substrates in general.
Typical examples include
benzene, water and noble gases adsorbed on various transition metal
surfaces,~\cite{kelkkanen2011,vanin2010,toyoda2010,wellendorff2010,abad2011,liu2012b,lee2012,yildirim2013,hamada2010,kumagai2011,poissier2011,tonigold2012,nadler2012,carrasco2012,silva2003,chen2011,victor2012,ambrosetti2012}
C$_{60}$ on Au(111),~\cite{hamada2011}
graphene on Ni(111) and Ir(111),~\cite{busse2011,mittendorfer2011,li2012,kozlov2012}
self-assembled monolayers of thiolates on Pt(111),~\cite{addato2012} 
pyridine on Cu(110),~\cite{atodiresei2009}
and isophorone on Pd(111).~\cite{liu2012}
A common conclusion from these studies is that the inclusion of vdW forces into DFT-GGA calculations
often results in a large increase in binding energies and in adsorption distances that are in better agreement with experimental
data. Moreover, in some instances vdW forces can also influence qualitatively the adsorption mechanism,
allowing for different minima in binding curves,~\cite{mittendorfer2011,li2012}
promoting chemisorption,~\cite{atodiresei2009} tipping the balance between chemisorption and physisorption 
states,~\cite{liu2013b} or enabling specific reaction pathways.~\cite{liu2012} 

In previous papers,~\cite{liu2012b,liu2013} we showed that when vdW interactions
are accurately accounted for, quantitative treatment of both chemisorbed and physisorbed benzene molecules
on metal surfaces becomes possible. In addition, Yildrim {\it et al.} have recently applied a range of van der Waals
density functionals (vdW-DF) to study the binding of benzene on metal surfaces, underscoring this
conclusion.~\cite{yildirim2013} However, despite this recent progress, many important
questions remain. For example, there are major gaps in our understanding of the general strengths and 
limitations of the various vdW-corrected methods when applied to these systems. Also very little is known about 
the performance of these methods at non-equilibrium adsorbate-substrate distances. From the chemical
perspective our understanding of the physical nature of the interaction between benzene and metal 
surfaces is still quite shallow as is our understanding of which metal surfaces can support physisorbed
(so-called precursor) states of benzene. In the current follow-on paper we extend our work on these 
systems aiming precisely to address these important unresolved issues. 
To this end we have considered two of the most
recently developed approaches: the
PBE+vdW$^{\rm surf}$ method~\cite{victor2012} and vdW-DF~\cite{dion2004} together with some
of its offspring.~\cite{murray2009,lee2010,klimes2010} For comparison
we have also included some selected calculations using the Perdew-Burke-Ernzerhof (PBE)
exchange-correlation functional~\cite{perdew1996} and the PBE+vdW scheme.~\cite{tkatchenko2009}

We have applied these 
approaches to  the adsorption of benzene on the (111) surface of Cu, Ag, Au, Rh, Pd, Ir, and Pt. The 
benzene-metal system is of interest in basic and applied
surface science~\cite{tautz2007,gomez2001,jenkins2009,MRS-Kronik,tan2005} and is an excellent 
model system to test different vdW-inclusive methods because it has been extensively
studied both theoretically~\cite{sautet1991,sautet1994,sautet1996,morin2003,morin2004,morin2006,kelkkanen2011,vanin2010,toyoda2010,wellendorff2010,abad2011,liu2012b}
and experimentally~\cite{zhou1990,wander1991,weiss1993,xi1994,stranick1995,syomin2001,ihm2004,mantooth2007,abad2009,toyoda2009,abad2011}
in recent years.
In particular, we used PBE+vdW$^{\rm surf}$ and vdW-DF approaches to obtain adsorption energies
and equilibrium geometries which we compare with the most recent
experimental data available. From the current study we find that both PBE+vdW$^{\rm surf}$ and certain improved 
versions of vdW density functionals are capable of reaching quantitative agreement with experimental measurements.
In addition, we computed the binding-energy curves of both physisorbed and chemisorbed 
benzene molecules on all the investigated metal surfaces. In our previous work we found that vdW interactions
are important to properly describe a metastable precursor state on Pt and Ir surfaces.~\cite{liu2012b}
Here we also extend this study to the chemisorption of benzene on Rh and Pd, where interestingly we find that
such metastable precursor states are not present. This leads us to the suggestion that only sufficiently polarizable metal surfaces are 
able to give rise to physisorbed precursor states. We also notice that different choices for including relativistic effects
lead to some deviations in the binding-energy curves.

This paper is organized as follows. Section \ref{sec:methods} sets out the details of the DFT calculations and 
approaches considered. Section \ref{sec:results} reports and discusses computed adsorption energies,
equilibrium energies, and binding-energy curves obtained with different vdW-inclusive DFT approaches.
Section \ref{sec:conclusions} brings a recapitulation and summary of results.

\section{THEORETICAL METHODS AND COMPUTATIONAL DETAILS}\label{sec:methods}

\subsection{PBE, PBE+vdW, and PBE+vdW$^{\rm surf}$ calculations \label{sec:methods1}}
 
We performed DFT calculations using the semi-local PBE
exchange-correlation functional~\cite{perdew1996} and two different vdW-inclusive approaches:
the PBE+vdW~\cite{tkatchenko2009} and the PBE+vdW$^{\rm surf}$ methods~\cite{victor2012}. 
The PBE+vdW method is based on a pairwise atom-atom approximation, whereas the 
PBE+vdW$^{\rm surf}$ goes beyond this approach and includes 
electrodynamic screening of vdW interactions 
by combining intermolecular vdW interactions with the Lifshitz-Zaremba-Kohn
theory~\cite{zaremba1976} for the dielectric screening within the metal surface.
PBE, PBE+vdW, and PBE+vdW$^{\rm surf}$ calculations were carried out using the
numeric atom-centered basis set all-electron code FHI-aims~\cite{blum2009,havu2009} applying
an atomic zeroth-order regular approximation (ZORA) for treating relativistic effects, in order to consistently compare with pseudopotential results.
We used ``tight'' settings, including the ``tier2'' standard basis set in the FHI-aims code for H and C
atoms, and the ``tier1'' basis set for transition metal atoms. We set the following 
thresholds for the convergence criteria: 
0.01 AA$^{-1}$  for the final forces in all structural relaxations,
10$^{-5}$ electrons for the electron density, and
10$^{-4}$ eV for the total energy of the system.  A 
Monkhorst-Pack~\cite{monkhorst1976} grid with 18$\times$18$\times$1 {\bf k}-point sampling per (1$\times$1) unit cell was used
for slab calculations.
We applied a 
dipole correction along the direction perpendicular to the metal
surface.~\cite{makov1995,neugebauer1992} 
These computational settings guarantee 
a convergence in the adsorption energies and equilibrium distances better than
0.01 eV and 0.01 \AA, respectively (Table \ref{tab:convergence}).

The unreconstructed close-packed (111) surfaces were modeled
by periodic (3$\times$3) unit cells, containing 6 atomic layers separated by at
least 20 \AA\ of vacuum, which ensures that the interaction between the adsorbed benzene
molecule and the periodic images of the metal slab is negligible. The 
benzene molecule and the uppermost two metal layers were allowed to 
relax during geometry relaxation. The four bottom metal layers were fixed at their
bulk-truncated positions using the lattice constants of the bulk metals from each
method (Table \ref{tab:parcel}).  Although the PBE+vdW$^{\rm surf}$
method slightly overestimates the interaction between metallic electrons in 
the bulk as discussed in Ref. [\cite{liu2013}], the performance of the PBE+vdW$^{\rm surf}$
method for lattice constants is very similar to PBE.

\subsection{Non-local density functional calculations \label{sec:methods2}}

We considered the following non-local exchange-correlation functionals:
the vdW-DF of Dion {\it et al.},~\cite{dion2004}
the vdW-DF2 of Murray {\it et al.}~\cite{murray2009} and Lee {\it et al.};~\cite{lee2010}
and three optimized versions of the vdW-DF.~\cite{klimes2010,klimes2011}
The vdW-DF2 functional aims to improve the 
binding description around energy minima in relation to vdW-DF
by changing both the exchange and non-local correlation components. 
In the  optimized versions of the vdW-DF, its original GGA functional
has been replaced by an optimized PBE (optPBE), optimized Becke88 (optB88), or 
optimized Becke86b (optB86b) to improve the accuracy
of both vdW-DF and vdW-DF2 
schemes. These three optimized functionals are
referred to as optPBE-vdW, optB88-vdW, and optB86b-vdW herein, respectively.


All vdW density functional calculations 
were carried out self-consistently within the VASP code~\cite{kresse1993,kresse1996}
as implemented by Klime\v{s} {\it et al.}~\cite{klimes2011} using the algorithm of 
Rom\'{a}n-P\'{e}rez and Soler.~\cite{roman-perez2009} In all cases the core 
electrons were replaced by PBE-based projector augmented wave (PAW)
potentials.~\cite{kresse1999}
We treated explicitly the  H (1$s$), C (2$s$, 2$p$), 
Cu (3$p$, 3$d$, 4$s$), Ag (4$d$, 5$s$), Au (5$d$, 6$s$), Rh (4$d$, 5$s$), Pd (4$d$, 5$s$),
Ir (5$d$, 6$s$), and Pt (5$d$, 6$s$) electrons as valence electrons and their
wavefunctions were expanded in plane-waves with a cut-off energy of 500 eV.
A Monkhorst-Pack~\cite{monkhorst1976} grid with 12$\times$12$\times$1
{\bf k}-point sampling per (1$\times$1) unit cell was used. We again applied a 
dipole correction along the direction perpendicular to the metal
surface~\cite{makov1995,neugebauer1992} and geometry optimizations 
were performed with a residual force threshold of 0.03 eV\AA$^{-1}$.

Using the lattice constants of the bulk metal from each
functional (Table \ref{tab:parcel}), we built up 6-layer slabs with a (3$\times$3)
unit cells and separated by at least 12 \AA\ (18 \AA\ when computing binding curves).
The metal atoms in the three bottom layers were fixed at their bulk-truncated 
positions during structure relaxation.
The metal lattice constants computed here (Table \ref{tab:parcel})
are in good agreement (differences are less than 0.020 \AA) with the values reported in
the literature using the same vdW functionals.~\cite{klimes2011,yildirim2013}
In general, vdW-DF and 
vdW-DF2 tend to overestimate all values with respect to experimental data. The largest
errors are 0.191 \AA\ (0.268 \AA) for vdW-DF (vdW-DF2) in the case of Ag (Au). This 
behavior can be traced back to the fact that these two functionals are too 
repulsive at short interatomic separations,~\cite{murray2009} which is important for
the correct description of lattice constants. The use of less repulsive exchange functionals
at short separations reduces the errors significantly,~\cite{klimes2011} giving better agreement with 
the reference experimental data for the transition metals considered here. Specifically,
the largest differences with experimental values are reduced to 0.113 \AA, 0.093 \AA,
and 0.052 \AA\ by using optPBE-vdW, optB88-vdW, and optB86b-vdW, respectively. Overall
the performance of these functionals, and in particular optB88-vdW and optB86b-vdW, for 
transition metal lattice constants is as good or better than PBE and PBE+vdW$^{\rm surf}$.

\subsection{Adsorption energies and geometries \label{sec:methods3}}

The adsorption energies of benzene on the investigated (111) metal surfaces 
were computed as follow:

\begin{equation}\label{adene}
	E_{\rm ads} = E_{\rm Bz-M} - E_{\rm M} - E_{\rm Bz},
\end{equation}

\noindent where $E_{\rm Bz-M}$ is the total energy of the adsorbed benzene 
molecule, $E_{\rm M}$ is the total energy of the relaxed bare metal slab, and 
$E_{\rm Bz}$ is the total energy of a relaxed gas-phase benzene molecule 
in the same unit cell used to compute the total system, but without the metal
slab. In addition, we considered the non-local correlation part, $E^{\rm nlc}$,
of the total exchange-correlation energy to compute the corresponding
non-local correlation contribution to the adsorption energy,
$E_{\rm ads}^{\rm nlc}$, as follow:

\begin{equation}\label{adnlc}
	E_{\rm ads}^{\rm nlc} = E_{\rm Bz-M}^{\rm nlc} - E_{\rm M}^{\rm nlc} - E_{\rm Bz}^{\rm nlc}.
\end{equation}

\noindent 
where the subindexes Bz-M, M, and Bz stand for the adsorbed benzene molecule,
the relaxed bare metal slab, and a relaxed gas-pahse benzene molecule. 
 
Equilibrium geometries were characterized by considering two average 
perpendicular heights ($d_{\rm CM}$ and $d_{\rm HM}$ for C--metal and H--metal
distances, respectively), which are referenced to the average $z$ positions of the 
relaxed topmost metal atoms.

We computed the adsorption energy of the system as a function of the 
perpendicular height, $d$, between the carbon backbone of the benzene 
molecule and the metal surface. We evaluated $d$ relative to the position 
of the topmost metal layer of the bare surface.
Notice that due to the optimization
constraints imposed on the carbon backbone (the $z$ coordinate of the 6 carbon atoms
was fixed) of the benzene molecule, the location of 
equilibrium minima in the computed binding curves may slightly differ in some cases
from the fully relaxed calculations described above.

\subsection{Level of convergence \label{sec:methods4}}

The computational settings described in Sections \ref{sec:methods1} and \ref{sec:methods2} 
guarantee a tight convergence in adsorption energies and equilibrium distances  as assessed
by an extensive series of convergence tests
performed for the adsorption of benzene on Au and Pt. Table \ref{tab:convergence}
summarizes the computed $E_{\rm ads}$, $d_{\rm CM}$, and $d_{\rm HM}$ values using
PBE+vdW$^{\rm surf}$ and optB88-vdW  as a function of the basis set size, 
{\bf k}-point mesh size, vacuum between slabs, total number of layers in the slab, and number of 
relaxed layers in the slab. In particular, PBE+vdW$^{\rm surf}$ (optB88-vdW) benzene adsorption
energies are better than 0.02 eV (0.02 eV) on Au and better than
0.04 eV (0.01 eV) on Pt. Equilibrium distances are in all cases better than 0.01 \AA.

Aiming at comparison between FHI-aims and VASP results, we computed the adsorption energy
of benzene on Au and Pt using PBE with these two codes. The FHI-aims (VASP) results for Au and 
Pt are $-$0.08 eV(-0.03 eV) and $-$1.00 eV ($-$1.02 eV), respectively. These differences are very
small and mainly reflect differences in the underlying basis sets: all-electron plus  atomic ZORA treatment of
relativistic effects (FHI-aims) and plane-waves plus PAW potentials (VASP).

\section{RESULTS AND DISCUSSION}\label{sec:results}

\subsection{Adsorption energies and equilibrium geometries}

\begin{figure}[htb]
\includegraphics[width=0.49 \textwidth]{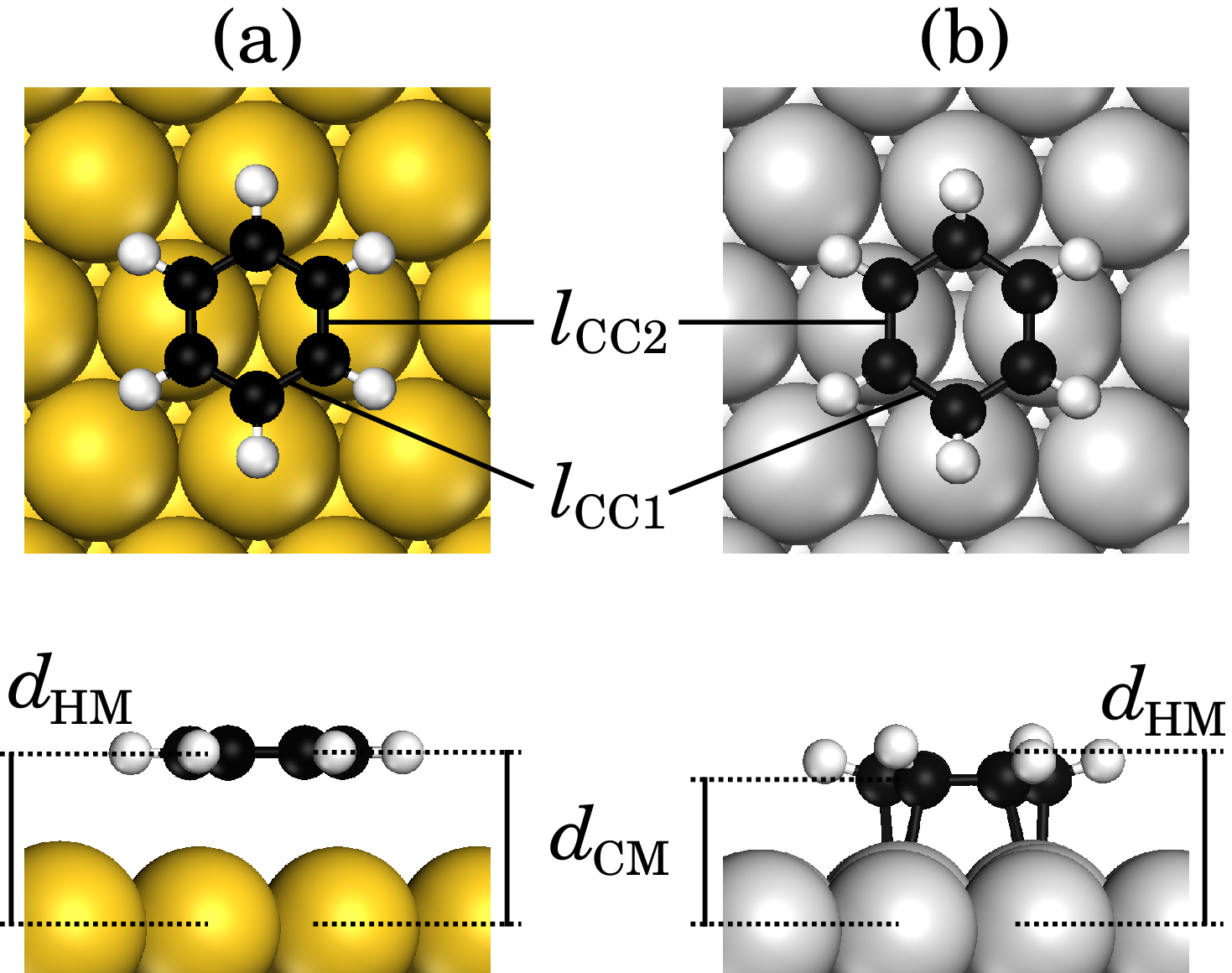}
\caption[]{Top and side views of a benzene molecule adsorbed at the bri30$^\circ$ site on the (111)
surfaces of Cu, Ag, and Au (a) and Pd, Pt, Rh, and Ir (b). The average heights between carbon (hydrogen)
and metal surface atoms, $d_{\rm CM}$ ($d_{\rm HM}$), and the average C--C bond lengths
within the adsorbed benzene molecule, $l_{\rm CC1}$ and $l_{\rm CC2}$,  are indicated.
The  $d_{\rm CM}$ ($d_{\rm HM}$) height is defined as the perpendicular distance between 
the two $xy$ parallel planes containing the average $z$ coordinate of the C (H) atoms within the benzene molecule
and the average $z$ coordinate of the metal atoms in the topmost surface layer. 
\label{fig:1}}
\end{figure}

Benzene adsorption is weaker on (111) coinage metal surfaces (Cu, Ag, and Au) than on
transition metals with unfilled $d$ bands (Rh, Pd, Ir, and Pt).~\cite{jenkins2009} 
For example, the experimental adsorption energy of benzene on Au is
0.73--0.87 eV,~\cite{syomin2001} whilst the interaction with Pt is stronger,
1.57--1.91 eV.~\cite{ihm2004} This is a direct consequence of the position of the $d$-band
center, which is substantially below the Fermi level in the case of coinage metals. 
A weak benzene-metal interaction (physisorption) implies a flat potential energy
surface (PES) and, therefore, benzene molecules can easily diffuse over the surface, as 
it has been found by scanning tunneling microscopy (STM) experiments on Cu(111) and
Au(111) terraces at low temperatures.~\cite{abad2009,mantooth2007,stranick1995}
These observations have been supported recently by DFT calculations by some of us,~\cite{liu2012b,liu2013}
where the predicted energy difference among eight high-symmetry adsorption sites of benzene on
Au(111)~\cite{saeys2002} was  only 0.04 eV using PBE+vdW$^{\rm surf}$.
Among the most preferable adsorption sites was a bridge
configuration with an angle of 30$^\circ$ 
between the C--C and Au--Au bonds (referred to as bri30$^\circ$ herein). This adsorption geometry is
shown in  Fig. \ref{fig:1}. The preference for the bri30$^\circ$ site in the case of Pt
is also supported by low-energy electron diffraction (LEED)~\cite{wander1991}
and STM~\cite{weiss1993} experiments.

In our previous work \cite{liu2013} we showed that the bri30$^\circ$ site is preferred for
chemisorption (on Rh, Pd, Ir, and Pt) of benzene when considering PBE, PBE+vdW, and PBE+vdW$^{\rm surf}$.
In the case of physisorbed systems (on Cu, Ag, and Au) the relative stability among
most of the eight high-symmetry adsorption sites is very small (less than 0.02 eV), but all three
methods predict the bri30$^\circ$ site to be in the group of the most favorable adsorption
sites. In the following we focus on the bri30$^\circ$ site to examine the performance of a
range of different methods:
PBE+vdW, PBE+vdW$^{\rm surf}$, vdW-DF, vdW-DF2, optPBE-vdW, optB88-vdW,
optB86b-vdW, and for reference PBE.
The computed adsorption energies and equilibrium geometries of benzene on
Cu, Ag, Au, Rh, Pd, Ir, and Pt are summarized in
Tables \ref{tab:ads-ene} and \ref{tab:ads-geom}, respectively.
A key observation is that PBE yields negligible adsorption energies of benzene on coinage metal
surfaces and largely underbinds on Rh, Pd, Ir, and Pt. For example,
the PBE adsorption energy on Pt is $ca.$ 0.9 eV lower than the experimental value
(Table \ref{tab:ads-ene}). The adsorption energies are systematically improved in all cases
upon inclusion of vdW forces by means of the PBE+vdW method. Further refinement of these 
calculated values is achieved when explicitly accounting for  
the dielectric screening by electrons inside the bulk metal through the PBE+vdW$^{\rm surf}$ method,
which takes into account the reduction of both the $C_6$ coefficient and vdW radius of the metal
atoms.~\cite{liu2013} 
In particular, the effect of screening decreases the adsorption energies on coinage
metals and Rh by 0.1--0.2 eV, whereas the interaction is enhanced in the case of
Pd, Ir, and Pt. It is notable that the inclusion of the
collective response effects reduces both vdW $C_6$ coefficients and vdW
radii, leading to opposite effects in the vdW energy and resulting in
non-trivial behaviour for different metals.~\cite{liu2013}

\begin{figure}[htb]
\includegraphics[width=0.99 \textwidth]{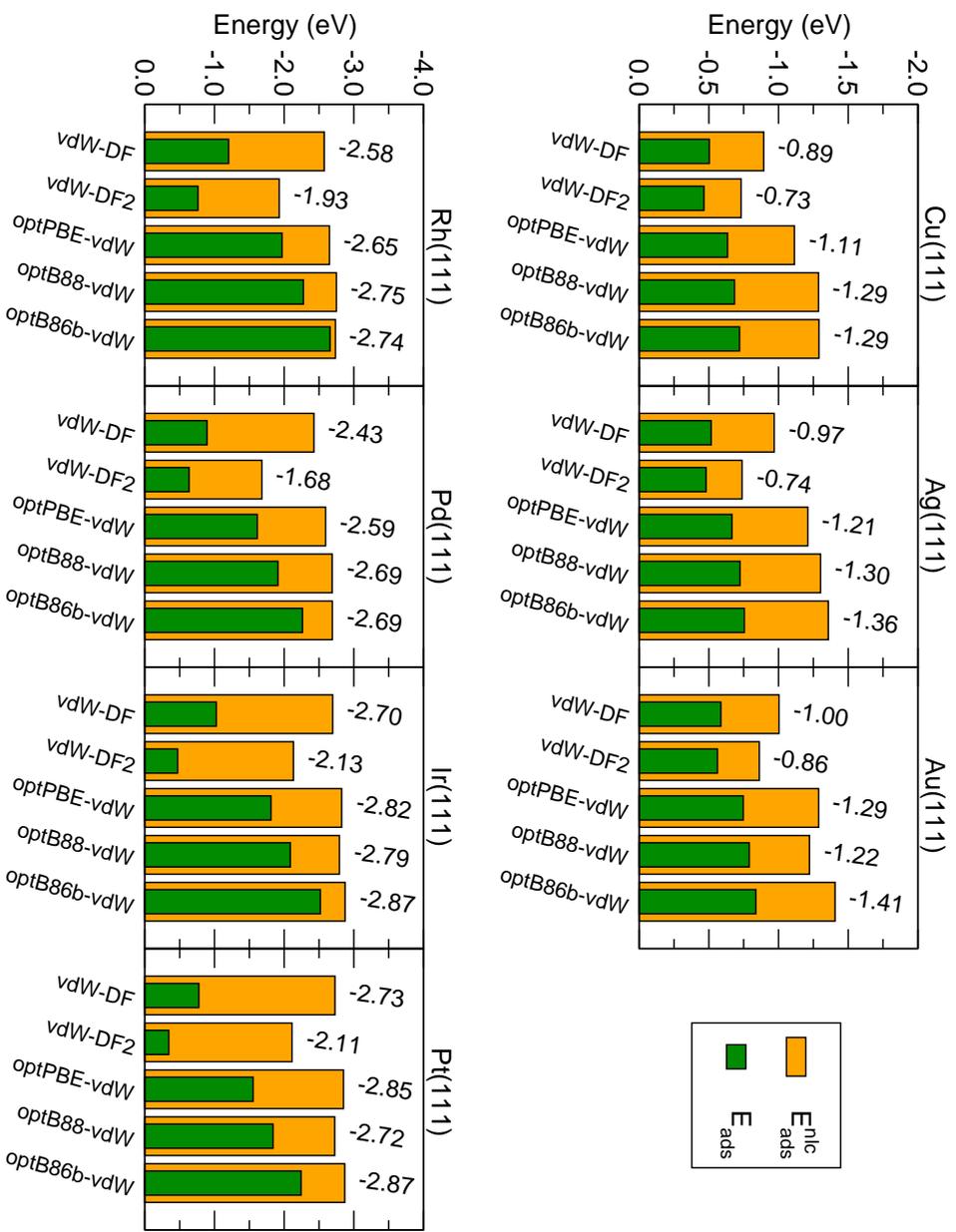}
\caption[]{Adsorption energies (thin centered green bars) and non-local correlation contribution to
those (thick orange bars and associated numbers) of a benzene molecule adsorbed on the (111)
surfaces of Cu, Ag, Au, Pd, Pt, Rh, and Ir. Five different density functionals are considered:
vdW-DF, vdW-DF2, optPBE-vdW, optB88-vdW, and optB86b-vdW.

\label{fig:2}}
\end{figure}

Moving to the results from the vdW density functionals, we find that the agreement with the available
experimental data strongly depends on the particular choice of the underlying exchange functional.
Consistent with previous studies for various adsorption
systems, the vdW-DF and vdW-DF2 functionals yield even smaller adsorption energies than
PBE.~\cite{dion2004,klimes2010,puzder2006,ziambaras2007,lee2010,hamada2010b,liu2012b,carrasco2012} 
Nevertheless, switching to improved
underlying exchange functionals (optPBE, optB88, and optB86b) consistently provides
larger adsorption energies than PBE and  quantitative agreement (especially for optB88-vdW) with
PBE+vdW$^{\rm surf}$ and available experimental data.
Similar findings have recently been reported by Yildirim {\it et al.}~\cite{yildirim2013}
It is important to stress that for the S22 database all the optimized density functionals considered 
here essentially yield similar results.~\cite{klimes2010} In contrast, for benzene adsorption 
on transition metals, especially in the case of strongly bound systems, the situation is more complex.
In particular, adsorption energies differ up to 0.6-0.7 eV on the four reactive metals when using optPBE-vdW,
optB88-vdW, or optB86b-vdW (Table \ref{tab:ads-ene}).
Further analysis shows that adsorption energy enhancements to the benzene-metal bonding are
largely due to non-local correlation, $E_{\rm ads}^{\rm nlc}$ (Eq. \ref{adnlc}),
which is the principal attractive contribution to the total
adsorption energy at the equilibrium geometry (Fig. \ref{fig:2}). Indeed the magnitude of
$E_{\rm ads}^{\rm nlc}$ on all surfaces is greater than the total adsorption energy.
On the most reactive metals (Rh, Pd, Ir, and Pt) the larger covalent character
of the bonding (chemisorption) is reflected by larger 
$E_{\rm ads}^{\rm nlc}$ than on coinage metals. 
We notice that vdW-DF2 systematically yields the lowest
$E_{\rm ads}^{\rm nlc}$ compared to the other functionals (especially in the case of the more reactive metals),
which explains why the smallest adsorption energies are found for this functional (Table \ref{tab:ads-ene}). 

The three optimized exchange functionals (optPBE, optB88, and optB86b) show similar  non-local
correlation contributions upon any given metal surface. Specifically,  the differences between 
$E_{\rm ads}^{\rm nlc}$ values computed using these three exchange functionals for a given system
are less than 0.18 eV on coinage metals and less than 0.10 eV on reactive metals (Fig. \ref{fig:2}).
This behavior explains why on the coinage metal series the adsorption energies computed with
optPBE-vdW, optB88-vdW, or optB86b-vdW are very similar: the energy differences are less than 0.10 eV
(Table \ref{tab:ads-ene}). On the other hand, on reactive metals the computed adsorption energies show
much larger energy differences despite $E_{\rm ads}^{\rm nlc}$ being similar.
This result highlights the fact that on such systems the benzene-metal bonding is dominated by covalency.
In particular, we found that the energy difference between $E_{\rm ads}^{\rm nlc}$ and $E_{\rm ads}$ provides a 
useful descriptor to rationalize the behavior of each functional when describing the 
benzene-metal bonding on chemisorbed systems. To this end, we define the ratio of attractive
non-local correlation lost after adding all the remaining contributions to the bonding,
$\chi_{\rm ads}^{\rm nlc}$, as

\begin{equation}\label{delta-nlc-ads}
	\chi_{\rm ads}^{\rm nlc} = \frac{E_{\rm ads}^{\rm nlc} - E_{\rm ads}}{E_{\rm ads}^{\rm nlc}}.
\end{equation}

\begin{figure}[t!]
\includegraphics[width=0.65 \textwidth]{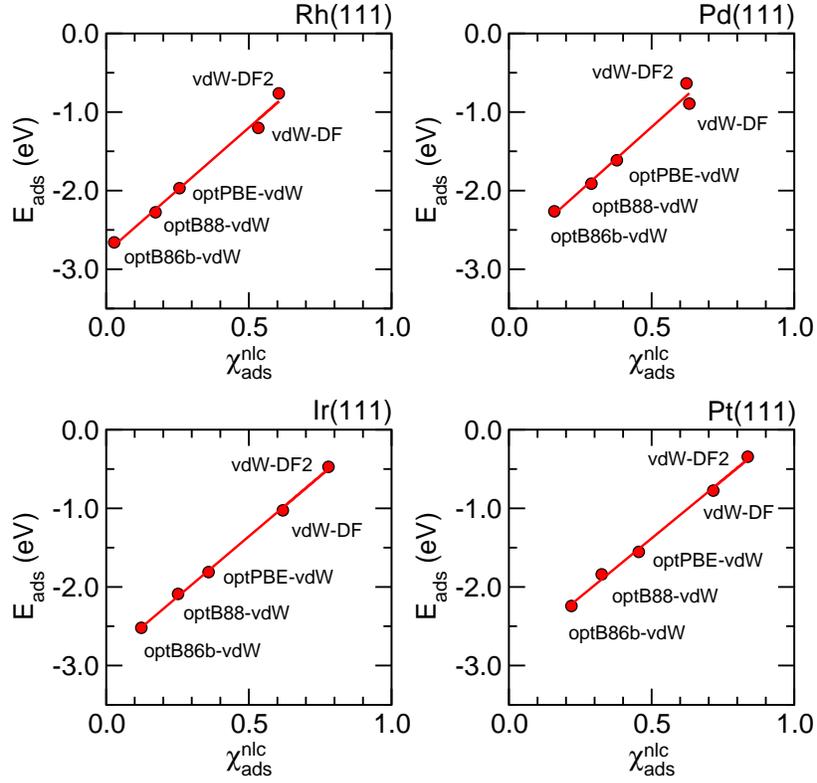}
\caption[]{Adsorption energy as a function of $\chi_{\rm ads}^{\rm nlc}$ (Eq. \ref{delta-nlc-ads}) for benzene
on Rh, Pd, Ir, and Pt using vdW-DF, vdW-DF2, optPBE-vdW, optB88-vdW, and optB86b-vdW.
The solid lines correspond to fitted linear regressions. 
\label{fig:3}}
\end{figure}

\noindent Essentially, $\chi_{\rm ads}^{\rm nlc}$ indirectly quantifies how repulsive a given 
functional is. This is important because the position of the repulsive Pauli wall is the main aspect that
ultimately controls the performance of a vdW density functional at short distances (equilibrium distances
in chemisorption). As shown in Fig. \ref{fig:3}, we find a linear correlation between $E_{\rm ads}$ and
$\chi_{\rm ads}^{\rm nlc}$ for all chemisorbed systems. 
The adsorption
energy decreases when increasing $\chi_{\rm ads}^{\rm nlc}$, i.e. a more repulsive exchange functional 
yields lower adsorption energies.
Interestingly, vdW-DF2 also fits well on this linear regression, even though it has a different underlying 
correlation term than the rest of density functionals.
From Fig. \ref{fig:3} it is clear that vdW-DF2 is the most repulsive
functional followed by vdW-DF, optPBE-vdW, optB88-vdW, and optB86b-vdW.
All three optimized exchange functionals show $\chi_{\rm ads}^{\rm nlc}$ values below 0.5, 
whereas vdW-DF and vdW-DF2 lead to larger values. Considering that optB88-vdW shows the best 
agreement  with available experimental data, Fig. \ref{fig:3} suggests that  density functionals with
$\chi_{\rm ads}^{\rm nlc}$ values ranging between 0.2 and 0.4 are particularly suitable to achieve 
a good description of the adsorption energy and equilibrium geometry for these systems.
Therefore, $\chi_{\rm ads}^{\rm nlc}$ of adsorbed benzene on late transition metals could be
a good descriptor to assess the accuracy of a given vdW-DF type density functional for 
other adsorbates on late transition metals. Actually similar conclusions are also 
found for water adsorption on transition metals.~\cite{carrasco2012} Nevertheless, further work is required
to see whether this is a general trend.

Another key observation is that on Rh, Pd, Ir, and Pt the benzene loses its gas-phase
planar configuration upon adsorption with hydrogen atoms tilting upward after relaxation and resulting
in two averaged perpendicular heights: a relatively short C--metal ($d_{\rm CM}$) and
a relatively long H--metal ($d_{\rm HM}$) distances as shown in 
Fig. \ref{fig:1}b and Table \ref{tab:ads-geom}. In contrast, on coinage metal surfaces the benzene molecule
adsorbs almost flat (Fig. \ref{fig:1}a). This qualitative difference in the equilibrium geometry between chemisorbed and
physisorbed systems is captured by all the methods considered, including PBE. From a quantitative
viewpoint, the performance of each method is less homogeneous though. Focusing first on physisorbed
systems, including the vdW energy through the PBE+vdW method has a dramatic impact on $d_{\rm CM}$,
which is shortened by 0.41--0.70 \AA\  with respect to PBE distances (Table \ref{tab:ads-geom}). 
Collective screening effects bring the benzene molecule even closer to the the surface by
0.18--0.25 \AA\ (cf. PBE+vdW and  PBE+vdW$^{\rm surf}$ values in Table \ref{tab:ads-geom}). 
In this respect,  vdW-DF predicts very large $d_{\rm CM}$, even larger in most cases than the PBE values.
Although vdW-DF2 substantially improves the vdW-DF equilibrium geometries, the 
differences with PBE+vdW$^{\rm surf}$ are still large, 0.24--0.56 \AA\ (Table \ref{tab:ads-geom}).
The other density functionals predict equilibrium geometries close to the
PBE+vdW$^{\rm surf}$ values, in particular optB88-vdW and optB86b-vdW present $\Delta d_{\rm CM}$
equal to 0.16--0.29 \AA\ and 0.07--0.33 \AA, respectively (Table \ref{tab:ads-geom}). Considering 
now the chemisorbed case, relatively similar equilibrium distances are obtained with all eight methods
(Table \ref{tab:ads-geom}). This result indicates that in such strongly bound systems the chemical bond
largely controls the adsorption height. 

All investigated methods predict very similar internal structures of the benzene molecule.  In particular,
the C--C distances ($l_{\rm CC1}$ and $l_{\rm CC2}$)  are almost identical no matter whether
PBE, PBE+vdW, PBE+vdW$^{\rm surf}$, vdW-DF, vdW-DF2, optPBE-vdW, optB88-vdW, or optB86b-vdW
is used (Table \ref{tab:ads-geom}). This indicates that the chemical bonds within the benzene
molecule are similarly well described by all the methods considered.

\subsection{Binding-energy curves}

\begin{figure}[t!]
\includegraphics[width=0.59 \textwidth]{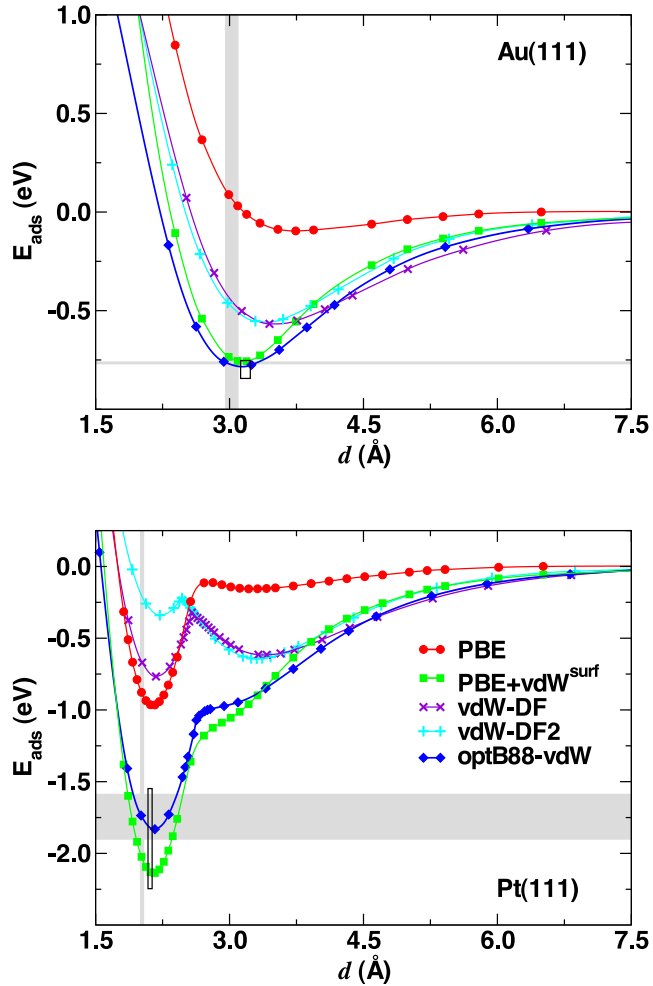}
\caption[]{Adsorption energy of benzene on Au(111) and Pt(111) as a function of the height $d$ (see text). Five
different methods are considered: PBE, PBE+vdW, PBE+vdW$^{\rm surf}$, vdW-DF, vdW-DF2, and optB88-vdW.
The lines are merely a guide to the eye. Here, PBE+vdW$^{\rm surf}$ calculations are carried out using
atomic ZORA for describing scalar relativistic effects. Our previous calculations in Refs. [\cite{liu2012b,liu2013b,liu2013}] 
employed scaled ZORA, leading to better agreement with experiments and to the appearance of a shallow 
physisorption precursor state for benzene on Pt.  The  grey bands indicate experimental binding distance (vertical) and
adsorption energy (horizontal) ranges. The open black rectangles indicate the range of 
binding distances and adsorption energies from optPBE-vdW and optB86b-vdW as given in
Tables \ref{tab:ads-ene} and \ref{tab:ads-geom}: in the case of benzene adsorption on Pt, different 
optimized density functionals yield adsorption energy differences up to 0.7 eV. This situation is in contrast 
to the S22 database, for which all the optimized non-local functionals considered here essentially yield 
similar results.
\label{fig:4}}
\end{figure}

We discuss now the role of vdW forces on the the binding-energy curves and their dependence on the 
specific method used. First we calculated the PBE, PBE+vdW$^{\rm surf}$, vdW-DF, vdW-DF2, and
optB88-vdW binding curves of a benzene molecule on Au and Pt as representatives of
physisorption and chemisorption, respectively (Fig. \ref{fig:4}). Each point in these graphs was obtained by 
keeping fixed the carbon backbone height $d$ from the surface as described in Sec. \ref{sec:methods3}.
On the two metal surfaces, of the methods considered, PBE predicts the weakest interaction at long range
(for $d >$ 6.0 \AA), pointing out its lack of non-local vdW  forces.
All the other methods recover to some extent
the vdW-attraction at long range.
Important quantitative
differences among the vdW-inclusive approaches considered remain though.
For example, vdW-DF and vdW-DF2
present a general tendency toward larger binding energies than PBE on Au and the 
physisorption region of Pt. The situation is significantly worse for
the chemisorption region on Pt, where the two functionals
predict the chemisorbed well to be shallower than the PBE minimum. Indeed vdW-DF2 predicts the 
chemisorbed well to be even shallower than the 
precursor physisorption-well minimum. This situation is much improved when considering optB88-vdW for both
Au and Pt,  optB88-vdW predicts a well depth and potential minimum location that lie close to the experimental data. Similarly,
PBE+vdW$^{\rm surf}$ shows good performance.   
Both optB88-vdW and PBE+vdW$^{\rm surf}$
give similar estimates for equilibrium distances and binding energies, especially in the case of Au. In addition,
the precursor physisorption state for
Pt  is only clearly  well defined when using vdW-DF and vdW-DF2, whilst PBE+vdW$^{\rm surf}$ and optB88-vdW
hardly predict a true minimum, but a barrier-less transition between the chemisorbed and physisorbed states.
We note in passing that the PBE+vdW$^{\rm{surf}}$ calculations
 in this paper employ atomic ZORA for the treatment of scalar relativistic effects to enable one-to-one comparison
 with pseudopotential calculations. Our previous work with scaled
ZORA,~\cite{liu2012b,liu2013b,liu2013} which explicitly includes scalar-relativistic effects in the orbital energies of the
full system, yielded better agreement with experiments and lead to the appearance of a shallow physisorption
precursor state for benzene on Pt.

\begin{figure}[t!]
\includegraphics[width=0.99 \textwidth]{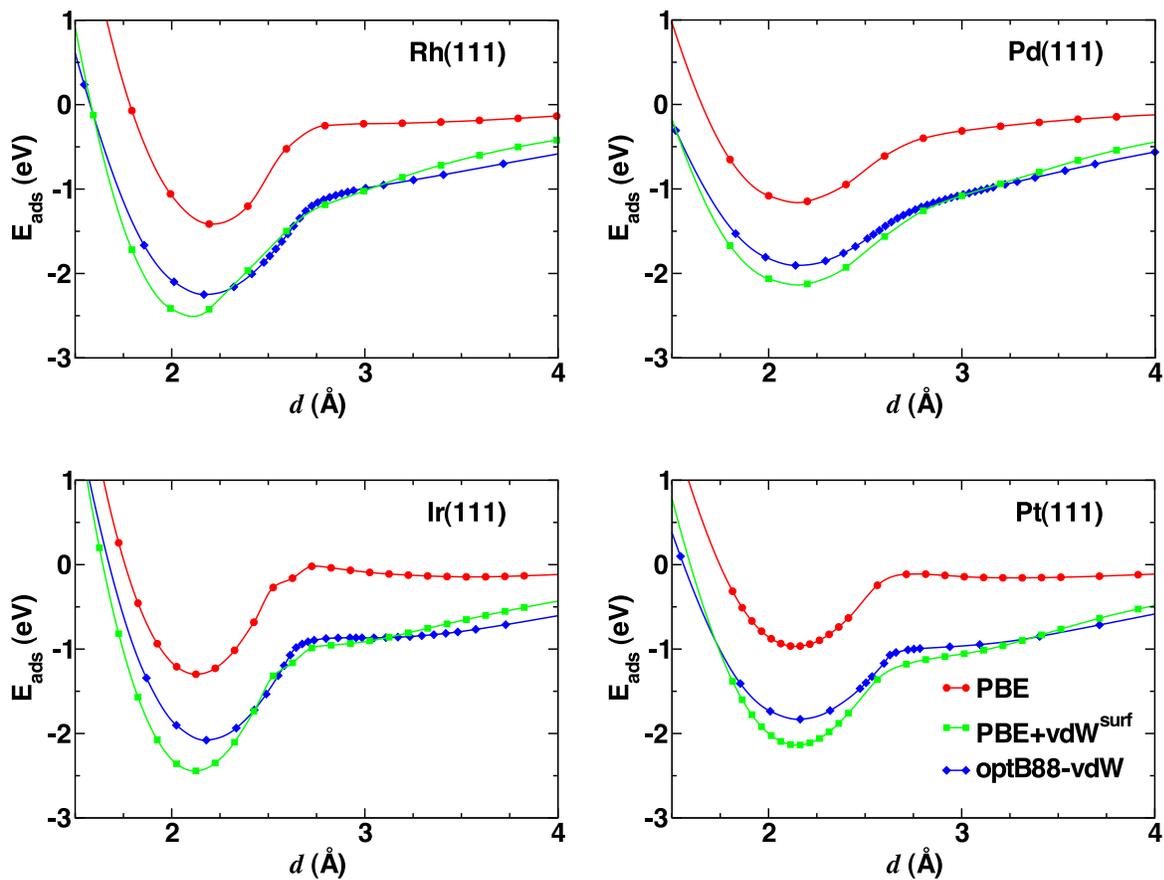}
\caption[]{Adsorption energy of benzene on Rh(111), Pd(111), Ir(111), and Pt(111) as a function of the height
$d$ (see text). Three different methods are considered: PBE, PBE+vdW$^{\rm surf}$, and optB88-vdW. The lines are
merely a guide to the eye. 
\label{fig:5}}
\end{figure}

We notice the existence of a very shallow metastable precursor physisorption state on Ir(111)
when considering optB88-vdW (Fig. \ref{fig:5}), where the barrier to adsorption predicted by PBE essentially vanishes.
Nevertheless, on Rh and Pd there exists a perturbation to the binding curve that suggests an underlying small physisorption-like interaction 
between 3.0-3.5 \AA, but neither PBE+vdW$^{\rm surf}$ nor optB88-vdW are able to resolve a clear minimum.
Since the stability of the physisorption state is determined mainly by vdW interactions,~\cite{liu2012} this result
suggests that the larger polarizability of the 5$d$ (Ir and Pt) compared to the 4$d$ (Rh and Pd) metals for the
same number of valence electrons is responsible for this behavior. In particular, the $C_6$ coefficients computed
with PBE+vdW$^{\rm surf}$ are $C_6^{Ir}$=98 hartree $\cdot$ bohr$^6$ to be compared with
$C_6^{Rh}$=84 hartree $\cdot$ bohr$^6$ and  $C_6^{Pt}$=120 hartree $\cdot$ bohr$^6$ to be compared with
$C_6^{Pd}$=102 hartree $\cdot$ bohr$^6$.

Another important difference between the binding-energy curves in Fig. \ref{fig:4} is the description
of the interaction at large distances from the surface, where the vdW attraction dominates. In this case
different methods give rise to rather different asymptotic decays. For example, when comparing 
vdW-DF and vdW-DF2, both binding curves lie very close to each other in the Pauli-repulsion
region at short binding distances on Au and the physisorption region on Pt, but vdW-DF2 decays faster
in the vdW-attraction region at separations larger than 4 \AA. 
In order to quantify these differences we have considered the traditional picture of physisorption where
the long-ranged vdW attractive potential
goes approximately like (see, e.g., Refs. [\cite{zaremba1976,bruch2007,bruch2009}])

\begin{equation}\label{eq:fit}
	V_{\rm vdW}(d)=C_3(d - d_0)^{-3}.
\end{equation}

\noindent Here $d$ is the distance of the adsorbate normal to the 
uppermost surface layer of atom cores in the solid. Usually one can fit the adsorption energy as a function
of $d$ by optimizing this expression \cite{dasilva2005,ambrosetti2012b} and  estimating the strength
of the asymptotic vdW attraction, $C_3$, and the surface image reference-plane, $d_0$.
Often $d_0$ is very close to one half of the metal interlayer separation ($\sim$1.2 \AA\ for Au and Pt).
Following this procedure we have fitted the data shown in Fig. \ref{fig:4} for PBE+vdW$^{\rm surf}$ and
$d$ values larger than 9 \AA. Only the vdW contribution to the adsorption energy in this region was taken into
account for the fitting. Table \ref{tab:c3} shows the corresponding  $C_3$ coefficients for 
Au and Pt. We note that the PBE+vdW$^{\rm surf}$ energy expression is explicitly constructed to reproduce
the \emph{exact} $C_3$ coefficient by employing the experimental dielectric function and accurate
polarizabilities for atoms in molecules. Following a similar procedure we fitted the $C_3$ coefficients for the vdW-DF type
density functionals. In this case we considered the non-local correlation contribution to the adsorption
energy (Eq. \ref{adnlc}) and we fixed $d_0$ to the values obtained with PBE-vdW$^{\rm surf}$ (1.06 \AA\ and 
1.03 \AA\ for Au and Pt, respectively). This procedure ensures a proper comparison between the two 
types of vdW-inclusive methods. As shown in Table \ref{tab:c3}, the $C_3$ coefficients depend on the method. 
The vdW-DF2 shows the weakest vdW interaction at this long distance range, yielding $C_3$
coefficients nearly two times smaller  than the rest of methods. This is consistent with the fact 
that vdW-DF2 heavily underestimate the $C_6$ coefficients of molecules.~\cite{vydrov2010}
On the other hand vdW-DF and optB88-vdW (both sharing the same non-local correlation energy expression by construction)
yield $C_3$ coefficients in good agreement with PBE+vdW$^{\rm surf}$. We note that such
good agreement might be partially accidental, since the vdW-DF correlation
energy yields errors of $\sim$20\% in the intermolecular $C_6$ coefficients.~\cite{vydrov2010}
However, the agreement in $C_3$ coefficients between PBE+vdW$^{\rm surf}$ and vdW-DF (optB88-vdW)
further explains the good performance of these methods found for the equilibrium properties of
benzene adsorbed on transition metal surfaces.

\section{CONCLUSIONS}\label{sec:conclusions}

We have investigated the performance of a series of vdW-inclusive DFT schemes
(PBE+vdW, PBE+vdW$^{\rm surf}$, vdW-DF, vdW-DF2, optPBE-vdW, optB88-vdW, and optB86b-vdW)
to compute the adsorption of benzene molecules on transition metal (111) surfaces.
Our calculations have demonstrated
that an accurate treatment of vdW dispersion interactions is essential to properly account for adsorption 
energies and equilibrium distances not only in weakly bound systems (Cu, Ag, and Au), but also in strongly
bound ones (Rh, Pd, Ir, and Pt). In particular, the performance of PBE+vdW$^{\rm surf}$ and  optB88-vdW for predicting
adsorption energies and equilibrium geometries are similar and both methods are in agreement with available
experimental data.

The performance of the different density functionals analyzed is sensitive to their underlying  exchange 
functional. 
We put forward that only optimized exchange functionals (optPBE, optB88, and optB86b) are capable of providing reliable results.
In this regard, a proper description of the short-ranged Pauli repulsion is critical to accurately
describe the investigated chemisorbed systems. 
We found a convenient descriptor to assess the 
magnitude of such short-ranged repulsion given a particular 
vdW-DF type density functional, namely, the energy difference ratio between the total adsorption energy and its non-local
correlation contribution, $\chi_{\rm ads}^{\rm nlc}$ (Eq. \ref{delta-nlc-ads}). A large $\chi_{\rm ads}^{\rm nlc}$ implies a more repulsive
density functional. We found that the poor performance of vdW-DF and vdW-DF2
can be tracked back to too large $\chi_{\rm ads}^{\rm nlc}$ values, whereas optimized exchange functionals 
(optPBE, optB88, and optB86b) are much less repulsive, presenting smaller $\chi_{\rm ads}^{\rm nlc}$ values and
superior accuracy. In particular, computed adsorption energies show a linear dependence with respect to
$\chi_{\rm ads}^{\rm nlc}$ and, therefore, this descriptor could be valuable for further exchange-correlation
functional development.
 
In addition,
optB88-vdW predicted a very shallow metastable precursor physisorption
state for Ir and Pt. For Rh, and Pd, the metastable state essentially vanished using either PBE+vdW$^{\rm surf}$ or optB88-vdW.
Both the
polarizability of molecules and substrates are key factors in the
appearance of precursor states.~\cite{liu2013b} Therefore, these results can be
explained by the lower polarizability of Rh versus Ir and Pd versus Pt.

Overall, our findings for benzene adsorbed on transition metals suggest that recent 
developments in vdW-inclusive methods have now reached  a sufficient state of maturity that
vdW interactions and many-body screening effects can be accounted for
at equilibrium distances for the specific adsorption systems considered.
Interestingly, PBE+vdW$^{\rm surf}$ and  optB88-vdW, which are two fundamentally different approaches,
are both able to achieve nearly equivalent quantitative agreement in adsorption energies and equilibrium
distances. Nonetheless some discrepancies between the approaches arise at large adsorbate-substrate 
distances. Resolving these discrepancies will probably require 
the application of more sophisticated methods. In general, comparison of methods on these and other 
adsorption systems along with more accurate experimental
measurements (both in terms of structure and energetics) of benzene
adsorption on well-defined metal surfaces remain highly desirable.

\section*{ACKNOWLEDGMENTS}

We are grateful for support from the FP7 Marie Curie Actions of the European Commission, via the Initial Training Network
SMALL (MCITN-238804) and the Career Integration Grant (FP7-PEOPLE-2011-CIG \texttt{NanoWGS}).
J.C. is a Ram{\'o}n y Cajal fellow  and Newton Alumnus supported by the Spanish
Government and The Royal Society. W.L. was funded by a fellowship from the Alexander von Humboldt Foundation.
A.M. is supported by the ERC (ERC Starting Grant \texttt{QUANTUMCRASS}) and by The Royal Society through a
Wolfson Research Merit Award. A.T. acknowledges support from the European Research Council
(ERC Starting Grant \texttt{VDW-CMAT}). We are grateful for computer time to 
the London Centre for Nanotechnology, the Barcelona Supercomputing Center, and the UK's national high performance computing service 
HECToR (from which access was obtained via the UK's Material Chemistry Consortium, EP/F067496).


\newpage

\newpage

\newpage

\newpage

\newpage

\newpage

\begin{table}[t!]
\centering
\caption{\label{tab:convergence} Adsorption energies in eV and equilibrium distances in \AA\ as
a function computational set-up for benzene adsorption on Au(111) and Pt(111) using
PBE+vdW$^{\rm surf}$ (FHI-aims) and optB88-vdW (VASP).}
\begin{tabular}{cccccccccccccccccccc}
\hline
\hline
&&& \multicolumn{7}{c}{Au} &&&& \multicolumn{7}{c}{Pt}\\
\hline
&&&  $E_{\rm ads}$ &&& $d_{\rm CM}$ &&& $d_{\rm HM}$ &&&& $E_{\rm ads}$ &&& $d_{\rm CM}$ &&& $d_{\rm HM}$\\
\hline
&&& \multicolumn{17}{c}{PBE+vdW$^{\rm surf}$}\\
\hline
{\bf Standard setup}\footnotemark[1] &&& {\bf$-$0.73} &&& {\bf 3.05} &&& {\bf 3.04} &&&& {\bf $-$2.14} &&& {\bf 2.08} &&& {\bf 2.51}\\
Tight tier3 &&& $-$0.72 &&& 3.05 &&& 3.04 &&&& $-$2.15 &&& 2.08 &&& 2.51\\
 8$\times$8$\times$1 {\bf k}-point mesh&&& $-$0.75 &&& 3.05 &&& 3.04 &&&& $-$2.19 &&& 2.08 &&& 2.51\\
30 \AA\ of vacuum&&& $-$0.72 &&& 3.05 &&& 3.04 &&&& $-$2.19 &&& 2.08 &&& 2.51\\
9 atomic layers slab &&& $-$0.74 &&& 3.05 &&& 3.04 &&&& $-$2.15 &&& 2.08 &&& 2.51\\
3 topmost atomic layers relaxed &&& $-$0.72 &&& 3.05 &&& 3.04 &&&& $-$2.14 &&& 2.09 &&& 2.52\\
\hline
&&& \multicolumn{17}{c}{optB88-vdW}\\
\hline
{\bf Standard setup}\footnotemark[2] &&& {\bf$-$0.79} &&& {\bf 3.23} &&& {\bf 3.23} &&&& {\bf $-$1.84} &&& {\bf 2.12} &&& {\bf 2.53}\\
Cutoff energy of 1000 eV &&& $-$0.79 &&& 3.23 &&& 3.23 &&&& $-$1.83 &&& 2.12 &&& 2.53\\
6$\times$6$\times$1 {\bf k}-point mesh &&& $-$0.78 &&& 3.23 &&& 3.23 &&&& $-$1.85 &&& 2.12 &&& 2.53\\
22 \AA\ of vacuum &&& $-$0.77 &&& 3.23 &&& 3.23 &&&& $-$1.83 &&& 2.12 &&& 2.53\\
9 atomic layers slab &&& $-$0.78 &&& 3.23 &&& 3.23 &&&& $-$1.83 &&& 2.12 &&& 2.54\\
4 topmost atomic layers relaxed &&& $-$0.79 &&& 3.23 &&& 3.23 &&&& $-$1.84 &&& 2.12 &&& 2.54\\

\hline
\hline
\end{tabular}
\footnotetext[1]{Tight tier2; 6$\times$6$\times$1 {\bf k}-point mesh; 20 \AA\ of vacuum; 6 atomic layers slab; 2 topmost atomic layers relaxed.}
\footnotetext[2]{Cutoff energy of 500 eV; 4$\times$4$\times$1 {\bf k}-point mesh; 12 \AA\ of vacuum; 6 atomic layers slab; 3 topmost atomic layers relaxed.}
\end{table}

\newpage

\begin{table}[htp]
\centering
\caption{\label{tab:parcel}Optimized lattice constants in \AA\ of bulk metals for different vdW-inclusive methods
considered in the present work. In addition, PBE values are shown. The values are compared to the 
experimental values corrected for the zero-point anharmonic expansion,~\cite{haas2009}
non-corrected values  are shown in parenthesis.}
\begin{tabular}{cc|ccccccc|cccccccc}
\hline
\hline
Method&&&Cu&&Ag&&Au&&&Rh&&Pd&&Ir&&Pt\\
\hline
PBE&&&3.631&&4.149&&4.159&&&3.830&&3.943&&3.871&&3.971\\
PBE+vdW&&&3.543&&4.071&&4.116&&&3.773&&3.913&&3.844&&3.939\\
PBE+vdW$^{\rm surf}$&&&3.572&&4.007&&4.163&&&3.765&&3.949&&3.873&&3.979\\
\hline
vdW-DF&&&3.699&&4.260&&4.241&&&3.877&&4.006&&3.929&&4.030\\
vdW-DF2&&&3.742&&4.329&&4.333&&&3.939&&4.075&&3.992&&4.107\\
optPBE-vdW&&&3.646&&4.181&&4.178&&&3.838&&3.951&&3.898&&3.989\\
optB88-vdW&&&3.623&&4.147&&4.158&&&3.826&&3.933&&3.891&&3.978\\
optB86b-vdW&&&3.596&&4.109&&4.119&&&3.800&&3.902&&3.866&&3.948\\
\hline
Experiment&&&3.596&&4.062&&4.062&&&3.793&&3.876&&3.831&&3.913\\
&&&(3.603)&&(4.069)&&(4.065)&&&(3.798)&&(3.881)&&(3.835)&&(3.916)\\
\hline
\hline
\end{tabular}
\end{table}

\newpage

\begin{table}[htp]
\centering
\caption{\label{tab:ads-ene}Calculated adsorption energies in eV for benzene on different
metal surfaces at the bri30$^\circ$ site. Available experimental data is also provided for comparison.
Experimental data for coinage metals has been
determined from TPD measurements and it has been corrected by the
method developed by Campbell and Seller~\cite{campbell2012} to reliably predict the
pre-exponential factor in the Redhead formula as described in Ref. [\cite{liu2013}].
}
\begin{tabular}{cc|ccccccc|cccccccc}
\hline
\hline
Method&&&Cu&&Ag&&Au&&&Rh&&Pd&&Ir&&Pt\\
\hline
PBE&&&$-$0.08&&$-$0.08&&$-$0.08&&&$-$1.44&&$-$1.16&&$-$1.26&&$-$1.00\\
PBE+vdW&&&$-$1.02&&$-$0.82&&$-$0.80&&&$-$2.73&&$-$2.02&&$-$2.35&&$-$1.98\\
PBE+vdW$^{\rm surf}$&&&$-$0.79&&$-$0.73&&$-$0.73&&&$-$2.46&&$-$2.13&&$-$2.40&&$-$2.14\\
\hline
vdW-DF&&&$-$0.50&&$-$0.52&&$-$0.59&&&$-$1.20&&$-$0.89&&$-$1.03&&$-$0.77\\
vdW-DF2&&&$-$0.47&&$-$0.48&&$-$0.56&&&$-$0.76&&$-$0.64&&$-$0.47&&$-$0.34\\
optPBE-vdW&&&$-$0.63&&$-$0.67&&$-$0.75&&&$-$1.97&&$-$1.61&&$-$1.81&&$-$1.55\\
optB88-vdW&&&$-$0.68&&$-$0.72&&$-$0.79&&&$-$2.27&&$-$1.91&&$-$2.09&&$-$1.84\\
optB86b-vdW&&&$-$0.72&&$-$0.76&&$-$0.84&&&$-$2.66&&$-$2.26&&$-$2.52&&$-$2.24\\
\hline
Experiment&&&$-$0.71\footnotemark[1]&&$-$0.69\footnotemark[2]&&$-$0.76\footnotemark[3]&&&&&&&&&$-$1.91--1.57\footnotemark[4]\\
\hline
\hline
\end{tabular}
\footnotetext[1]{Refs. [\cite{xi1994,campbell2012}] }
\footnotetext[2]{Refs. [\cite{zhou1990,campbell2012}].}
\footnotetext[3]{Refs. [\cite{syomin2001,campbell2012}].}
\footnotetext[4]{Microcalorimetry measurements from Ref. [\cite{ihm2004}].}
\end{table}

\newpage

\begin{table}[htp]
\centering
\caption{\label{tab:ads-geom}Calculated  $d_{\rm CM}$, $d_{\rm HM}$, $l_{\rm CC1}$, and $l_{\rm CC2}$
in \AA\  as defined in Fig. \ref{fig:1}. For coinage metal surfaces $l_{\rm CC1}$ has the same value as
$l_{\rm CC2}$.}
\begin{tabular}{cc|ccc|ccccccc|cccccccc}
\hline
\hline
&&&Method&&&Cu&&Ag&&Au&&&Rh&&Pd&&Ir&&Pt\\
\hline
&&&PBE&&&3.74&&3.69&&3.62\footnotemark[1]&&&2.14&&2.12&&2.15&&2.10\footnotemark[2]\\
&&&PBE+vdW&&&3.04&&3.14&&3.21\footnotemark[1]&&&2.14&&2.12&&2.14&&2.11\footnotemark[2]\\
&&&PBE+vdW$^{\rm surf}$&&&2.79&&2.96&&3.05\footnotemark[1]&&&2.12&&2.10&&2.13&&2.08\footnotemark[2]\\
$d_{\rm CM}$&&&vdW-DF&&&4.14&&3.95&&3.44\footnotemark[1]&&&2.19&&2.21&&219&&2.16\footnotemark[2]\\
&&&vdW-DF2&&&3.38&&3.40&&3.29\footnotemark[1]&&&2.25&&2.40&&2.23&&2.20\footnotemark[2]\\
&&&optPBE-vdW&&&3.28&&3.29&&3.22\footnotemark[1]&&&2.16&&2.15&&2.16&&2.12\footnotemark[2]\\
&&&optB88-vdW&&&3.08&&3.12&&3.23\footnotemark[1]&&&2.14&&2.13&&2.15&&2.12\footnotemark[2]\\
&&&optB86b-vdW&&&3.12&&3.10&&3.12\footnotemark[1]&&&2.13&&2.11&&2.13&&2.10\footnotemark[2]\\
\hline
&&&PBE&&&3.74&&3.70&&3.62&&&2.55&&2.47&&2.60&&2.54\\
&&&PBE+vdW&&&3.06&&3.15&&3.20&&&2.53&&2.47&&2.59&&2.53\\
&&&PBE+vdW$^{\rm surf}$&&&2.79&&2.95&&3.04&&&2.51&&2.46&&2.57&&2.51\\
$d_{\rm HM}$&&&vdW-DF&&&4.13&&3.95&&3.42&&&2.58&&2.53&&2.64&&2.57\\
&&&vdW-DF2&&&3.37&&3.39&&3.27&&&2.61&&2.60&&2.65&&2.65\\
&&&optPBE-vdW&&&3.27&&3.28&&3.22&&&2.56&&2.49&&2.59&&2.54\\
&&&optB88-vdW&&&3.06&&3.12&&3.23&&&2.53&&2.46&&2.59&&2.53\\
&&&optB86b-vdW&&&3.11&&3.10&&3.12&&&2.53&&2.46&&2.58&&2.52\\
\hline
&&&PBE&&&1.40&&1.40&&1.40&&&1.47 (1.44) &&1.45 (1.43)&&1.48 (1.44)&&1.47 (1.43)\\
&&&PBE+vdW&&&1.40&&1.40&&1.40&&&1.47 (1.44)&&1.45 (1.43)&&1.48 (1.44)&&1.48 (1.44)\\
&&&PBE+vdW$^{\rm surf}$&&&1.40&&1.40&&1.40&&&1.47 (1.43)&&1.45 (1.43)&&1.48 (1.44)&&1.48 (1.43)\\
$l_{\rm CC1}$&&&vdW-DF&&&1.40&&1.40&&1.40&&&1.47 (1.44)&&1.45 (1.43)&&1.48 (1.44)&&1.48 (1.44)\\
($l_{\rm CC2}$)&&&vdW-DF2&&&1.40&&1.40&&1.40&&&1.46 (1.43)&&1.43 (1.42)&&1.48 (1.43)&&1.47 (1.43)\\
&&&optPBE-vdW&&&1.40&&1.40&&1.40&&&1.47 (1.43)&&1.45 (1.43)&&1.48 (1.44)&&1.48 (1.44)\\
&&&optB88-vdW&&&1.40&&1.40&&1.40&&&1.46 (1.43)&&1.45 (1.43)&&1.48 (1.43)&&1.47 (1.43)\\
&&&optB86b-vdW&&&1.40&&1.40&&1.40&&&1.46 (1.43)&&1.45 (1.43)&&1.48 (1.44)&&1.48 (1.44)\\
\hline
\hline
\end{tabular}
\footnotetext[1]{To be compared with deduced data from work function and pentacene adsorption experiments,~\cite{abad2009,toyoda2009,abad2011} 3.03$\pm$0.08 \AA.}
\footnotetext[2]{To be compared with LEED measurements,~\cite{wander1991} 2.02$\pm$0.02 \AA.}
\end{table}

\newpage

\begin{table}[htp]
\centering
\caption{\label{tab:c3} $C_3$ coefficients in eV\AA$^3$
(obtained from Eq. \ref{eq:fit})
for benzene on Au and Pt
using PBE+vdW$^{\rm surf}$, vdW-DF, vdW-DF2, and optB88-vdW.
}
\begin{tabular}{cc|ccc|cc}
\hline
\hline
Method &&& Au &&& Pt\\
\hline
PBE+vdW$^{\rm surf}$&&& 9.16$\pm$0.08 &&& 9.77$\pm$0.13\\
vdW-DF &&& 8.94$\pm$0.18 &&& 9.76$\pm$0.20 \\
vdW-DF2 &&& 4.86$\pm$0.08 &&& 5.64$\pm$0.07  \\
optB88-vdW &&& 8.66$\pm$0.18 &&& 9.14$\pm$0.22\\

\hline
\hline
\end{tabular}
\end{table}


\end{document}